\documentstyle[epsfig,12pt]{article}
\begin{document}
\begin{flushright}
DTP-00/04\\
January 2000\\
\end{flushright}
\vspace*{7mm}
\begin{center}
{\Large{\bf{LOW ENERGY HADRON PHYSICS}}}\\
\vspace*{7mm} 
{\large{M.R. Pennington}}
\vspace*{0.5cm} \\
{ Centre for Particle Theory, University of Durham,
Durham DH1 3LE, U.K.}
\end{center}
\vspace{5mm}

\begin{abstract}
\baselineskip=4.5mm
\vspace*{4mm}
Ask a group of particle theorists about low energy hadron physics and they will say that this is a subject that belongs to the age of the dinosaurs. However, it is GeV physics that controls the outcome of every hadronic interaction at almost every energy.
Confinement of quarks and gluons (and any other super-constituents) means that it is the femto-universe that determines what experiments detect. What we still have to learn at the start of the 21st century is discussed.
\end{abstract}

\vspace*{3mm}
\baselineskip=6.mm
\parskip=1.5mm

\sloppy

\section{Low energy dynamics}

Low energy hadron dynamics is determined by the structure of the QCD vacuum.  Once we thought this ground state was empty, but now we know it is a seething cauldron of quarks, antiquarks and gluons.  Indeed, so strong are their interactions that they form all measure of condensates: $q{\overline q}$, ${\cal G}{\cal G}$, ${\overline q} {\cal G} q$, etc.  It is the nature of this vacuum and the scale of these condensates that determine low energy hadron physics. It is this that we can learn about at DA$\Phi$NE.

Most importantly, the vacuum determines the spectrum of hadrons and its scale.  Thus the nucleon  has a mass of 1 GeV, while conventional ${\overline q}q$ mesons, like the $\rho$, are two-thirds of this mass, but  pions merely 140 MeV.  Indeed, it is these masses squared that determines dynamics and having $m_{\pi}^2\,=\,0.02$ GeV$^2$ makes pions by far the lightest of all hadrons.  But why?  This is a question for which we have had a good idea of the answer for many decades, but in fact its only now that we are on the threshold of definitively testing.

We begin with QCD with two flavours of quark, for simplicity.  Then we have 
kinetic energy and interaction terms in the Lagrangian for both the up quark and the down.  Since the proton and neutron masses are almost equal, we can imagine that $m_u = m_d$.  Then QCD has an exact $SU(2)_F$ symmetry, which is reflected at the hadron level by the proton and neutron having the same strong interactions.  Now, the scale of QCD is fixed on renormalization by the momentum scale at which the strong interaction is strong.  This $\Lambda_{QCD} \sim 100-200$ MeV.  The current masses of the $u$ and $d$ quarks are very small. At a few MeV, they are much smaller than $\Lambda_{QCD}$ and so almost massless.  If we separate the quark wavefunction into left-handed and right-handed components, it is only the mass term in the Lagrangian that couples left and right.  So, if $m_u = m_d = 0$, there is no such coupling and we can make $SU(2)_F$ transformations in the right-handed space, quite independently of those on the left.  While this is a symmetry at the quark level, it is not there at the hadron level.  Scalar and pseudoscalars, vectors and axial vectors are not degenerate in mass.  Thus this symmetry must be broken at the hadron level.

\begin{figure}[t]
\begin{center}
~\epsfig{file=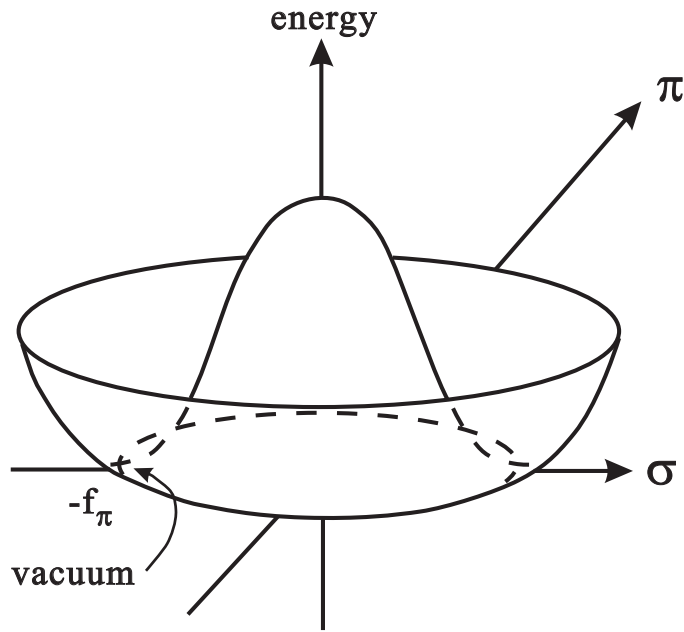,width=9.cm}
 \caption{\it The potential in the $\sigma-$model. With no explicit chiral 
symmetry breaking, there are degenerate minima all round the rim of the Mexican hat. In practice, the hat is tilted, so that there is a unique minimum close to the state labelled {\it vacuum}.}\end{center}
\vspace{-6mm}
\end{figure}

A way to picture this was introduced by Nambu~\cite{nambu} nearly forty years ago, long before the advent of QCD.  Consider a hadron world with just scalar, $\sigma$, and pion fields $\bf{\pi}$~\cite{gellmann}.  Then we can imagine that the form of the potential generated by their interactions, though symmetric in $\sigma$ and $\mid{\bf \pi}\mid$, may not be simply parabolic with a  minimum when these fields are zero, but could be Mexican hat shaped, Fig.~1.
Then the ground state chosen by nature is where the $\sigma$ field has a non-zero vacuum expectation value, Fig.~1.  The particles we observe  correspond to quantum fluctuations about this ground state. The scalar field is massive as the fluctuations go up and down the sides of the {\it hat}, while the Goldstone fluctuations are massless corresponding to motion round the {\it hat}.  This is an illustration of Goldstone's theorem with Nambu's ferromagnetic analogy.  It is natural to identify these Goldstone modes with the pions, or in a world with 3 massless quark flavours with the $\pi$, $K$ and $\eta_8$.

\begin{figure}[h]
\begin{center}
\epsfig{file=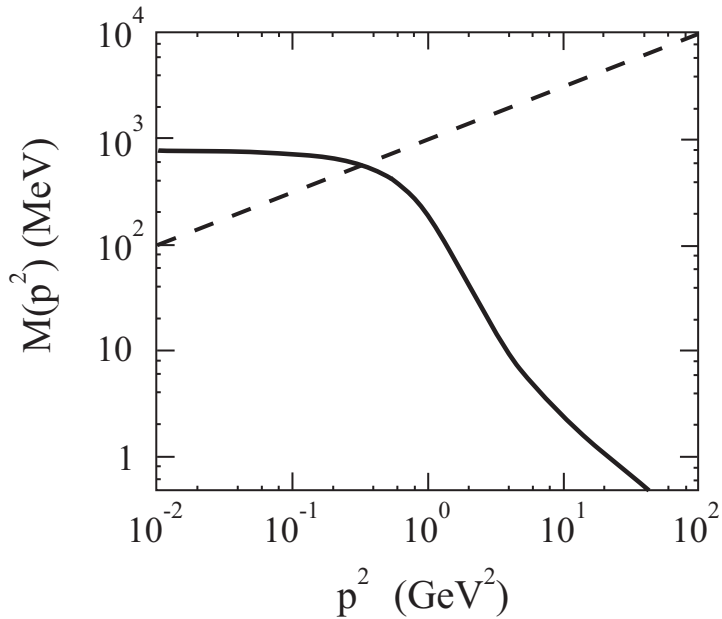,width=9.5cm}
 \caption{\it The mass function, $M(p^2)$, dynamically generated in QCD for a massless bare quark at Euclidean momenta~\cite{roberts}. The dashed line marks $M = p$; it's intersection with the solid line defines  the quark's ``euclidean'' mass.
     }
\end{center}
\vspace{-3mm}
\end{figure}

This is only a model of the hadron world~\cite{nambu,gellmann}. What does QCD tell us is happening at the quark-gluon level?  We can calculate the fully dressed quark propagator by summing all possible gluonic emissions and absorptions.  If the current mass of the quarks is zero, then it is well known that the fermion mass function remains zero at all orders in perturbation theory.  Thus any breaking of chiral symmetry must be non-perturbative.  This we might calculate on the lattice, but zero mass particles don't fit on a finite sized lattice.  Consequently, we most naturally calculate the quark propagator in the continuum using the Schwinger-Dyson equations.  These provide a genuinely non-perturbative set of integral equations in which the behaviour of the quark propagator is determined in terms of the dressed gluon (and ghost) propagators, and quark-gluon interactions~\cite{robertsrev,roberts}.  Such calculations have reached a stage of maturity that we can be certain that provided the effective quark-gluon coupling is large at infrared momenta less than $\Lambda_{QCD}$, then a non-zero mass function is generated even though the current mass is zero.  This mass function, Fig.~2~\cite{roberts}, produced by gluonic clouds and the $q{\overline q}$ sea is 350--400 MeV at low Euclidean momenta.  Thus a quark's constituent mass is dynamically generated.  Importantly, this is merely an {\it effective} low energy mass:  quarks being confined have no poles in their propagators and are never on-shell.  From the large momentum behaviour of this mass function
we learn this corresponds to the formation of $q{\overline q}$ condensates of size 
$$ \langle\; q{\overline q}\; \rangle_0\;=\;- (240\,{\rm MeV})^3\quad .$$
A value totally consistent with QCD sum-rule phenomenology.

\begin{figure}[h]
\begin{center}
~\epsfig{file=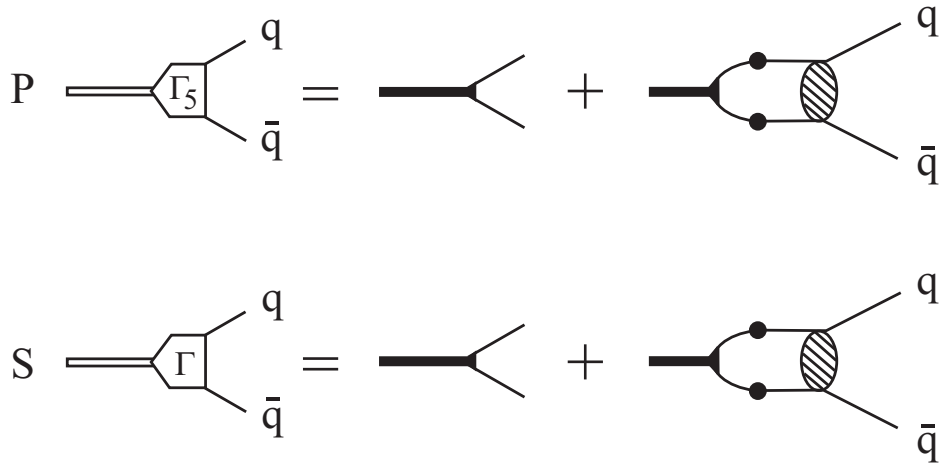,width=12.4cm}
 \caption{\it
     Bethe-Salpeter equation for a P(seudoscalar) and S(calar) $q{\overline q}$ bound state. The dots denote full quark propagators and vertices. The shaded 4-point function is the quark scattering kernel.
     }
\end{center}
\vspace{-3mm}
\end{figure}
With the light quark propagator determined, one can consider the $q{\overline q}$ bound states with pseudoscalar and scalar quantum numbers, Fig.~3.  The pseudoscalar bound state is massless, as one would expect for an explicit realisation of Goldstone's theorem.  In contrast, the scalar is massive, with a mass of roughly twice that of the dressed (or constituent) quark.  These scalars are the Higgs bosons of the strong interaction, responsible for the masses of all light hadrons~\cite{robertsrev,scadron}.

Though these QCD calculation and Nambu's modelling are quite consistent in their phenomenology, there has been the proposal of Jan Stern and collaborators~\cite{stern1} that chiral symmetry breaking need not be the result of such a large $q{\overline q}$ condensate.  Instead this condensate could be small, perhaps $(-100\,{\rm MeV})^3$,
 as in an antiferromagnetic analogy, and other condensates could be important in   breaking chiral symmetry.  Since physics is an experimental subject, we must test this hypothesis. The study of low energy $\pi\pi$ scattering provides just such an opportunity.
\begin{figure}[h]
\begin{center}
~\epsfig{file=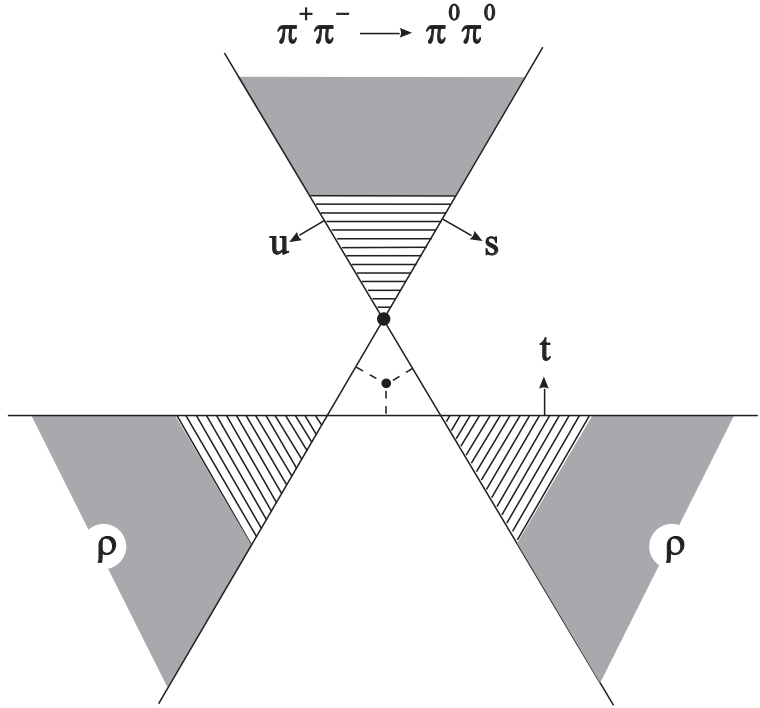,width=10.0cm}
 \caption{\it
 Mandelstam plane for $\pi\pi$ scattering. The symmetry point of the triangle and the $t-$channel threshold are marked. The shaded areas are those accessible in $K_{e4}$ decays. The grey regions are where measurements extracted from the one pion exchange component of $\pi N\to \pi\pi N$ start to apply through the $\rho-$resonance region. 
     }
\end{center}
\vspace{-5mm}
\end{figure}

Consider the amplitude ${\cal F}$ for the process $\pi^+\pi^-\to\pi^0\pi^0$ in the $t-$channel~\cite{daphce}. If one of the pions is massless, chiral symmetry  requires that in the limit when the 4--momentum of this massless pion goes to zero, the amplitude vanishes. Thus we have at the symmetry point of the Mandelstam triangle
${\cal F}(s=t=u=m_{\pi}^2)=0$ --- this is known as the Adler condition. This means that, though $\pi\pi$ scattering is by definition a strong interaction process, it is actually weak at low energies. It is then natural that we can make a Taylor series expansion of the low energy amplitude about the Adler point
in powers of momenta and the mass of the off-shell pion. The scale for this expansion is supplied by $m_{\rho}^2$ in hadron dynamics, or more naturally by
$32\pi f_{\pi}^2$ in chiral dynamics terms. Then at the symmetry point of the Mandelstam triangle with all the pions having their physical mass, we have the prediction
\begin{equation}
{\cal F}\left(s=t=u={\textstyle \frac{4}{3}} m_{\pi}^2\right)\,=\,\frac {\alpha\, m_{\pi}^2}{32\pi f_{\pi}^2}\,+\,\cdots\quad ,
\end{equation}
where the parameter $\alpha$ depends of the size of $\langle q\overline q \rangle$. In fact, $\alpha \simeq 1$ for Standard Chiral Perturbation Theory ($S\chi PT$)~\cite{gasserleutwyler} with a large condensate and $\alpha \simeq 4$ with a small condensate~\cite{stern2}, as is possible in Generalised $\chi PT$. Thus,
\begin{eqnarray}
{\cal F}\left(s=t=u={\textstyle \frac{4}{3}} m_{\pi}^2\right)&\simeq& 0.02\qquad\qquad {\rm in~~S\chi PT}\cr
&&\cr
{\cal F}\left(s=t=u={\textstyle \frac{4}{3}} m_{\pi}^2\right)&\le& 0.09\qquad\qquad {\rm in~~G\chi PT}\quad .
\end{eqnarray}
Though the size of the $\langle q{\overline q}\rangle-$condensate can engender a factor of 4 variation in the amplitude at the centre of the Mandelstam triangle, both versions of the Chiral Expansion agree in the physical regions, where for instance at the $\rho$ their parameters are fixed by experimental data, Fig.~4.
Thus only very close to threshold is there any testable difference between the large and small condensate predictions.

\begin{figure}[h]
\begin{center}
~\epsfig{file=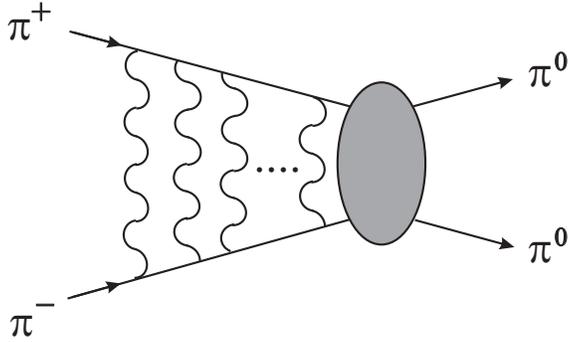,width=7.5cm}
 \caption{\it
Atomic pionium decays hadronically to $\pi^0\pi^0$.
     }
\end{center}
\vspace{-6mm}
\end{figure}
The closest to the symmetry point that one can get experimentally is by considering the lifetime of pionic atoms. This provides a direct measurement of the amplitude ${\cal F}$ at $t=4m_{\pi}^2, s=u=0$, Fig.~4. The decay rate, which is the inverse of the lifetime, is proportional to the modulus squared of this amplitude, with proportionality constants that have recently been corrected by higher orders in the electromagnetic coupling, Fig.~5, and for higher orders in the chiral expansion, among other considerations~\cite{bernatoms}. There is now a consensus on the relationship~\cite{hadatom}. PS212 at CERN, the DIRAC experiment~\cite{dirac}, is designed to measure the lifetime of such pionic atoms. DIRAC has had a successful run this year and the expectation is that with further running scheduled for 2000, they will be able to determine the scattering length
for $\pi^+\pi^-\to\pi^0\pi^0$ to an accuracy of 10\%, with the ultimate hope of $\pm 5\%$ in later years.

\begin{figure}[h]
\vspace{-2mm}
\begin{center}
~\epsfig{file=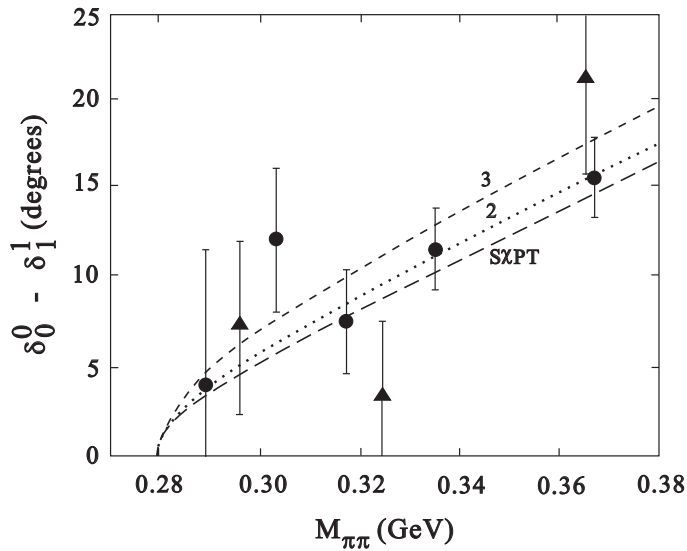,width=9.cm}
 \caption{\it
The phase difference $\delta^0_0 - \delta^1_1$ as a function of dipion mass.
The triangles and solid circles are from the Pennsylvania~\cite{penn} and Geneva-Saclay~\cite{genevasaclay} $K_{e4}-$decay
experiments, respectively. The curves are the predictions of two loop $\chi$PT: the Standard result for which $\alpha \simeq 1.2$ and for $\alpha = 2,3$ of Generalised $\chi$PT.
     }
\end{center}
\vspace{-7mm}
\end{figure}

Another experiment that we will hear about at this meeting involves $K_{e4}$ decays, both in E865 at Brookhaven~\cite{e865} and here with the KLOE detector~\cite{kloe}. With a branching ratio of $\sim 4. 10^{-5}$,
 kaons decay into a dilepton pair, mainly $e\nu$, and a charged dipion pair. Measuring the dependence of the decay distribution on its five kinematic variables, one can in principle determine the relative phase of the $\pi\pi$ system in an $S$ and $P-$wave.
By the famous theorem of final state interactions of Watson, this phase difference is exactly the same phase difference as in $\pi\pi$ scattering. The pions scatter universally, independently of the way they are produced.
The results for this phase difference from previous experiments by the University of Pennsylvania~\cite{penn}  with 7000 decays and by the Geneva-Saclay group~\cite{genevasaclay}
with 30000 are shown in Fig.~6, together with the predictions of S$\chi$PT
and for G$\chi$PT with $\alpha = 2$ and 3. One sees that the error bars from the current experiments have to be substantially smaller than those in Fig.~6 if we are going to be able  to
discriminate between values of $\alpha$ and hence the size of the $q{\overline q}-$condensate. However, the correlation with energy is just as important as individual data-points for this determination. This we will hear about too at this meeting~\cite{colangelo}.

\section{$\eta-\eta'$ mixing}

Another topic that addresses the nature of the QCD vacuum is the study of $\eta-\eta'$ mixing. We are all familiar with the mixing among the vector and tensor mesons, where the octet and singlet components mix ideally to give  physical states of definite quark flavour. We know the $\omega$ and $\phi$ are within $3^o$ of this ideal situation. The same pattern emerges whether one considers
the meson masses, their hadronic decays, their production in $e^+e^-$, etc. 
However, it has been known for some time that $\eta-\eta'$ mixing depends critically on the quantity considered~\cite{bramon,feldmann}. As recognised by Leutwyler~\cite{leutwyler}, two mixing angles are in fact required: one for the octet sector and the other for the singlet.
These differ because the singlet couples to the vacuum. 

First we specify decay constants $f_P^a$ for the nine pseudoscalars $P$ (with momentum $p$) by analogy with that of the pion from 
\begin{equation}
\langle\, 0 \mid A_{\mu}^a \mid P(p)\, \rangle \,=\, i \sqrt{2}\, f_P^a\, p_{\mu}\quad,
\end{equation}
where $a=0,1,...,8$. We can then define two mixing angles $\theta_0$ and $\theta_8$ by~\cite{leutwyler}
\begin{equation}
{f_{\eta}^8}/{f_{\eta'}^8}\,=\,\cot \theta_8\quad, \qquad 
{f_{\eta}^0}/{f_{\eta'}^0}\,=\,-\tan \theta_0 \quad .
\end{equation}
These angles are defined so that in the limit in which they vanish the $\eta_8 \to \eta$ and $\eta_0 \to \eta'$.
In the limit of  exact $SU(3)$ flavour, $\theta_0\,=\,\theta_8$.
Now, Leutwyler~\cite{leutwyler} has shown how one can expand the difference in these angles as
\begin{equation}
\sin\left(\theta_0 - \theta_8\right)\,=\, \frac{2\sqrt{2}\,\left(f_K^2 - f_{\pi}^2\right)}{4 f_K^2 - f_{\pi}^2}\;+\; \cdots\qquad .
\end{equation}
As the physical kaon decay constant $f_K \simeq 1.22 f_{\pi}$, this implies that $\theta_0 - \theta_8 \simeq 16^o$. We are thus far from the $SU(3)_F$ symmetry in this sector --- the singlet  couples to the vacuum.
In a recent comprehensive survey of $\eta-\eta'$ mixing, Feldmann~\cite{feldmann} has shown how the mixing in different situations for different quantities can all be reconciled by introducing these two mixing angles, which differ by $16^o \pm 2^o$.

The reason for the interest in the $\eta$ and $\eta'$ is because  of the way these states are intimately tied to the structure of the QCD vacuum. Clearly, the $\eta$ is only a little heavier than the kaons, but the $\eta'$ at 958 MeV is much heavier, particularly as it is the mass squareds that are relevant. In the $SU(3)$ chiral limit, when the masses of the $u$, $d$ and $s$ quarks all go to zero, the 8 flavour components of the octet axial vector current  of Eq.~(3) are all conserved and the $\eta_8$ is massless. However, the divergence of the singlet current is given by the topological charge density and according to the Adler-Bell-Jackiw anomaly is non-zero.
In contrast, in the limit of large $N_c$, all nine components of the divergence of the axial vector current vanish and the $\eta_0$ becomes a ninth Goldstone boson. Thus the mixing in this sector teaches us how close nature is to the chiral or to the large $N_c$ limits of QCD. 

\begin{figure}[h]
\begin{center}
\vspace{-11mm}
~\epsfig{file=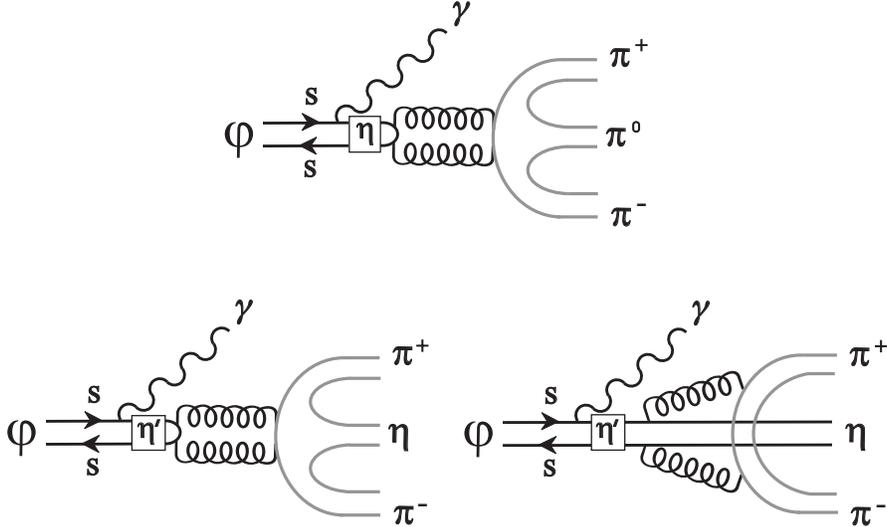,width=11.8cm}
 \caption{\it
Quark line graphs for $\phi$ radiative decay to $\eta$ and $\eta'$, which in turn decay to $3\pi$ and $\eta\pi\pi$, respectively.
     }
\end{center}
\vspace{-9mm}
\end{figure}
From the quark line diagrams shown in Fig.~7, it was recognised long ago by Deshpande and Eilam~\cite{deshpande}, 
and then by Rosner~\cite{rosner}, that $\phi$ radiatively decaying to the $\eta$ and $\eta'$ can test their strange quark content, the $\phi$ acting as an $s{\overline s}$ source. At that time the mixing was assumed to be simply governed by one mixing angle. Now we know better. Nevertheless, these decays provide a key test of the mixing, fixing parameters such as $\Lambda_3$, the coupling to the Wess-Zumino term~\cite{leutwyler,feldmann}. The branching ratio for $\phi\to\gamma\eta$ is reasonably well determined to be $(1.26 \pm 0.06)$\%~\cite{pdg}. From this, one predicts that
the corresponding ratio for $\phi\to\gamma\eta'$ is between 4 and 8 times $10^{-5}$ depending on the value of $-0.3\le \Lambda_3 \le 0.1$. The 1997 experimental value from CMD2~\cite{cmd} gave a far bigger range 
from the branching ratio of $\sim (12 \pm 7). 10^{-5}$. However, the more recent results from VEPP-2M~\cite{vepp} show that with much improved statistics, DA$\Phi$NE should be able not only to test our current understanding of $\eta-\eta'$ mixing, but to fix the parameters in such a scheme. 

\section{Scalar mesons}

The nature of the lightest scalar mesons has long been an enigma~\cite{mrp2}. There are far more than can fit into one quark model multiplet, Table~1. Some have claimed that there is a $\kappa$ at 900 MeV~\cite{kappa}, as well as at 1430 MeV, and so perhaps there are two nonets. The LASS data rules this out~\cite{LASS}, just as a re-analysis of the CERN-Munich polarised target data~\cite{lesniak} rules out the oft-claimed {\it narrow} $\sigma(750)$, see for example ~\cite{svec}. Even so, there are too many isoscalar states, Table~1. One may be a glueball, or several may be mixed with glue~\cite{glue,mrpglue}.
\begin{table}[h]
\vspace{-5mm}
\centering
\caption{ \it The light scalar mesons.
}
\vskip 0.1 in
\begin{tabular}{|l|c|c|} \hline
             &              &    \\
    Isospin     &  common name & PDG name \\
            &              &      \\
\hline
\hline
    ~~~~0   & $\sigma$  & $f_0(400-1200)$ \\
    & $S^*$     & $f_0(980)$ \\
     &           & $f_0(1370)$   \\ 
     &           & $f_0(1500)$   \\
     & $\theta$  & $f_0(1710)$  \\
~~~~$\frac{1}{2}$ & $\kappa$ & $K_0^*(1430)$ \\
 ~~~~1 & $\delta$ & $a_0(980)$ \\
   &        & $a_0(1450)$ \\

\hline
\end{tabular}
\label{extab}
\end{table}
\begin{figure}[t]
\begin{center} 
~\epsfig{file=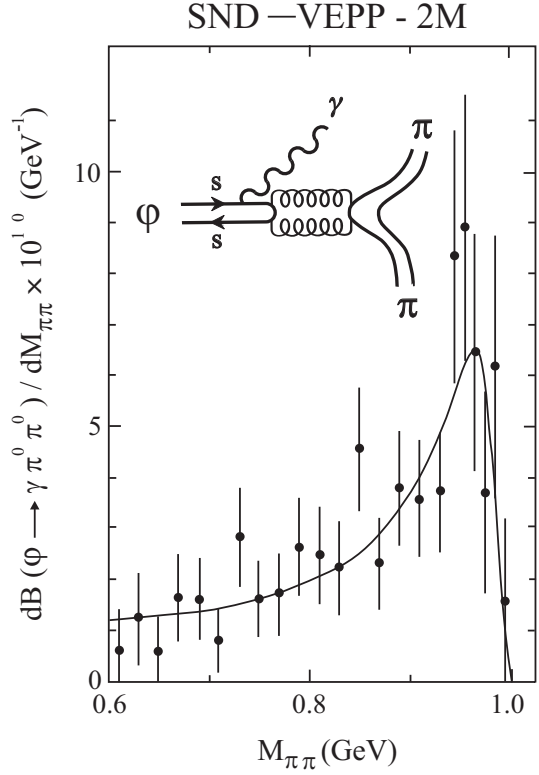,width=7cm}
 \caption{\it
Experimental results on radiative $\phi$ decay to $\pi^0\pi^0$ as measured by the SND group~\cite{snd} at VEPP-2M, together with a quark line representation of this process.
     }
\end{center}
\vspace{-3mm}
\end{figure}

A key aspect of the scalars is their dressing by hadrons. All mesons, of course,
spend part of their time in a multi-meson continuum, $\rho$ in $\pi\pi$, $\phi$ in $K{\overline K}$, etc., it is through these channels that these unstable particles decay. However, these hadronic dressings turn out to be a small perturbation on the underlying $q{\overline q}$ states. However, in the case of scalars, this is not the case. It is a phenomenological fact that their couplings to mesons are much larger.
Their lifetimes shorter. They are {\it looser} systems. Consequently, they readily form dimeson systems. The details may depend on the calculational scheme~\cite{tornqvist,vanbeveren,elena}. Nevertheless, even if the underlying scalar quark model multiplet is centred at 1400 MeV, the isotriplet $a_0$ and one of the isosinglet $f_0$ states automatically has such strong couplings
to $K{\overline K}$, that they spend an appreciable fraction of their lifetime
as $K{\overline K}$ systems that the masses of the physical hadrons are pulled to 1 GeV~\cite{tornqvist,elena}. That the $f_0(980)$ likes to couple strongly to $K{\overline K}$, or equivalently $s{\overline s}$ systems, is observed in its distinct appearance in channels with hidden strangeness. While the $f_0(980)$ appears as a sharp dip,
or a shoulder, in $\pi\pi\to\pi\pi$ and in central production, such as $pp\to pp\pi\pi$, it appears as a pronounced peak in $J/\psi\to\phi\pi\pi$, as observed long ago~\cite{mrp3}. This behaviour is reproduced by the $\pi\pi$ spectrum in $\phi$ radiative decay~\cite{snd} as shown in Fig.~8. Hopefully improved statistics at DA$\Phi$NE will sharpen up its signal.

Once again these mesons illuminate our knowledge of the QCD vacuum.
In the $1/N_c$ expansion, we expect processes in which flavour is annihilated
to be suppressed. This we recognise as the OZI-rule, Fig.~9.
\begin{figure}[b]
\begin{center} 
~\epsfig{file=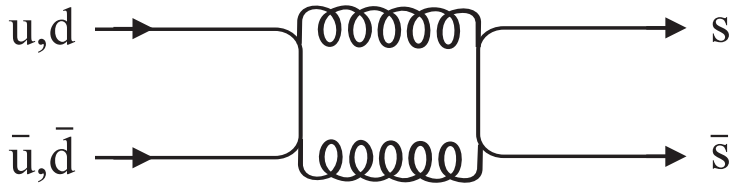,width=10cm}
 \caption{\it
Non-strange $q{\overline q}$ system coupling to $s{\overline s}$ would normally be suppressed by the OZI rule in the $1/N_c$ expansion.
     }
\end{center}
\vspace{-3mm}
\end{figure} 
Even though this
perturbative QCD argument is not expected to be as good for light flavours
with $N_c=3$, nevertheless within the large $N_c$ framework, this still applies.
However, the appearance of the $f_0(980)$  (and $a_0(980)$) mesons just at
$K{\overline K}$ threshold counters this suppression.
The OZI-rule not
merely fails, but is reversed. As emphasised by Moussallam~\cite{bachir}, this tests the flavour dependence of the QCD condensates. 

\begin{table}[h]
\vspace{-2mm}
\centering
\caption{ \it Predictions for the absolute rate for $\phi\to\gamma f_0(980)$ depending on the composition of the $f_0(980)$~\cite{closeachasov,brownclose}.
}
\vskip 0.1 in
\begin{tabular}{|l|c|} \hline
    Composition     &  BR$(\phi\to\gamma f_0(980))$ \\
\hline
\hline
 $qq{\overline{qq}}$ & ~$O(10^{-4})$  \\
 $s{\overline s}$   & ~$O(10^{-5})$  \\
 $K{\overline K}$    & $< O(10^{-5})$ \\ 

\hline
\end{tabular}
\end{table}

It has long been advertised that,  depending on the nature of these mesons, whether they are simple $q{\overline q}$, $4-$quark as $qq{\overline{qq}}$ or as $K{\overline K}-$molecule~\cite{closeachasov,brownclose,oset}, there are significant differences in the branching ratio for $\phi$ radiative decays to $f_0$ and $a_0$, see Tables~2, 3.
\begin{table}[h]
\vspace{-2mm}
\centering
\caption{ \it Ratio of width for $\phi\to\gamma a_0$ to that for $\phi\to\gamma f_0$ for different compositions of the scalar mesons~\cite{closeachasov,brownclose}. $n \equiv u, d$.
}
\vskip 0.1 in
\begin{tabular}{|l|c|} \hline
   Composition   &  $\Gamma(\phi\to\gamma a_0)/\Gamma(\phi\to\gamma f_0)$ \\
\hline
\hline
    $K{\overline K}$   &  1 \\
    $(n{\overline s})({\overline n}s)$  & 1 \\
    $(ns)(\overline {ns})$ &     9        \\ 
    $(n{\overline n})(s{\overline s})$ & structure dependent \\

\hline
\end{tabular}
\vspace{-2mm}
\end{table}
 In reality, the choices are not likely to be so stark. In the calculation of Tornqvist (and Boglione and myself)~\cite{tornqvist,elena} the physical $a_0(980)$ is only 20\% $q{\overline q}$ and 70\% $K{\overline K}$. An outstanding calculation is to deduce what this inevitable mixture means for the
predictions for their two photon decays and for their appearance in $\phi-$radiative decays. That is a challenge for theorists. 

\section{Challenge for DA$\Phi$NE}

The challenge for DA$\Phi$NE is to produce high statistics results on $K\to e\nu \pi^+\pi^-$, and on all of $\phi\to\gamma\eta$, $\gamma\eta'$, $\gamma f_0$ and $\gamma a_0$. Then we have the hope of understanding the nature of the QCD vacuum a little more clearly: a vacuum that shapes the world of hadrons,
of hadronisation and all other aspects of confinement --- an exciting prospect.

\section{Acknowledgements}
I would like to thank my many colleagues in the DA$\Phi$NE physics community for their enthusiasm for this project. I acknowledge travel support from the EEC-TMR Programme, Contract No. CT98-0169, EuroDA$\Phi$NE.

\newpage

\end{document}